\pdfoutput=1
\documentclass{my-paper}

\title{Emergent Outcomes of the veToken Model\thanks{This is a
preprint of a published article~\citep{lloyd-et-al-23}.}}

\author{Thomas~Lloyd\thanks{\email{tomlloyd1992@gmail.com}}}

\author{Daire~\'O~Broin\thanks{\email{daire.obroin@setu.ie},
\orcid{0000-0002-2886-5546}}}

\author{Martin~Harrigan\thanks{\email{martin.harrigan@setu.ie},
\orcid{0000-0002-6069-7001}}}

\affil{\gls*{setu}, \acrlong*{roi}}

\date{}

\begin{document}

\maketitle

\begin{abstract}
  Decentralised organisations use blockchains for governance: on-chain
transactions allocate voting weight, publish proposals, cast votes,
and enact the results.  A key challenge is aligning the short-term
outlook of pseudonymous voters with the long-term success of the
organisation.  The Vote-Escrowed Token (veToken) model attempts to
resolve this tension by requiring voters to lock tokens of value for
an extended period in exchange for voting weight.

In this paper we describe the veToken model and analyse its emergent
outcomes.  We describe its implementation by Curve, a popular
automated market maker for stablecoins, and the ecosystem of protocols
built on top.  We show that voting outcomes are strongly associated
with the bribes set by higher-level protocols, and that the cost per
vote varies depending on how it is acquired.  The outcomes of the
fortnightly votes held by Convex Finance closely track the
distribution of bribes through voting markets such as Votium.  Frax
Finance, a stablecoin issuer, plays a central role even though it
directly locks relatively few tokens with Curve; instead, it
indirectly locks tokens through yield aggregators and purchases voting
weight through voting markets.

Although the veToken model in isolation is straightforward, it leads
to complex and emergent outcomes.  Decentralised organisations should
consider these outcomes before adopting the model.

\end{abstract}

\section{Introduction}\label{sec:introduction}

Blockchains are a decentralising technology: they push
operations away from centralised entities and intermediaries and
toward individual actors and stakeholders.  Decentralised
organisations, or, the somewhat aspirationally named,
\glsstar{}{dao}{s}, involve stakeholders that manage shared resources
via a blockchain.  Frequently, they need to make decisions that
require the combined input of the stakeholders.  For example, a
decentralised finance platform such as Aave might want to decide on
supporting an additional
asset,\footnote{\url{https://app.aave.com/governance/proposal/168/}}
or an \glstext*{nft} collection such as Nouns \gls*{dao} might want to decide on
funding an animated short
movie.\footnote{\url{https://nouns.wtf/vote/195}} In the simplest
model, stakeholders are allocated voting weight using governance
tokens, where one token equals one vote.  They cast their votes using
on-chain transactions or off-chain signed messages.  This process
mirrors stockholder voting in corporate governance.

However, two significant characteristics of decentralised
organisations and the simple voting model are that stakeholders are
identified pseudonymously and voting weight (tokens) can be traded,
and even borrowed for short terms.  This creates a tension between the
desire of stakeholders for short-term gains and self-serving actions
and the long-term prospects of the organisation.  A particularly
egregious example of this tension occurred in April 2022 when a
malicious actor borrowed a controlling stake of tokens in the
Beanstalk \gls*{dao} and voted for a proposal to send
\dUSD[prefix=million~]{182} from the \glsstar{}{dao}{'s} treasury to
himself~\citep{immunefi-22}.  Although these actions resemble a
corporate raid, their speed and flagrancy were due to the pseudonymity
of the attacker and the borrowability of voting weight.

The \emph{vote-escrowed token} (veToken) model attempts to resolve
this tension by requiring stakeholders to \emph{lock} their tokens for
a fixed time period in exchange for voting weight.  Locked tokens
cannot be transferred.  Therefore, the voting preferences of the
stakeholders should better align with the long-term prospects of the
decentralised organisation.  In this paper we explain the veToken
model and analyse its emergent outcomes.  We describe the
implementation of the model by Curve, a popular automated market
maker, and the ecosystem of protocols that has arisen on top of it.
These include yield aggregators such as Convex Finance, voting markets
such as Votium, and significant participants such as Frax Finance, a
stablecoin issuer that is itself governed by a \gls*{dao}.
We show that the veToken model has created a complex ecosystem with
emergent outcomes.

Specifically, this paper examines the dual nature of the veToken
model.  The model avoids the shortcomings of earlier token-voting
designs by aligning stakeholder and organisational goals.  However,
this has led to the emergence of additional protocols keen on using
its distinctive features.  In the context of Decentralised Finance
(DeFi), such developments are regarded as beneficial and described as
\textquote[][.]{money legos}  Yet for governance, it creates a dynamic
in which influence is disproportionately exercised at the upper tiers
of the governance ecosystem, where the rules of engagement have been
distorted so as not to match those of the underlying organisation.
This distortion does not refer to the modification of the veToken
contract, as the contract is immutable, but to the alteration of the
intended incentive structure, lockup logic, and distribution of
governance influence through composable smart contracts that enable a
higher tier of governance.  It is not necessarily a violation of the
underlying protocol's rules.  We demonstrate this by examining the
impact of Votium's incentivisation mechanisms on the outcomes of
Convex's fortnightly proposals.  We then examine Frax, which locks
relatively few tokens directly with Curve but locks substantially more
through Convex and acquires further voting weight through Votium
bribes.

The paper is organised as follows.  We survey related work in
\cref{sec:related-work}, specifically, blockchain-based decentralised
governance and token-based voting.  \Cref{sec:vetoken-model} introduces
the veToken model and its implementation within the Curve ecosystem.
We explain the relevance of gauges (\cref{sec:gauges}), gauge
proposals, and higher-level protocols such as Convex, Votium, and Frax
(\cref{sec:higher-level-protocols}).  In \cref{sec:method} we describe
our data collection process.  \Cref{sec:results} is the core of the
paper.  We analyse the veToken model at the Curve, Convex, and Votium
levels.  We show that voting outcomes are strongly associated with
bribes (\cref{sec:votes-follow-bribes}) and that votes can be acquired at
different prices at the different levels
(\cref{sec:the-cost-of-votes}).  Finally we conclude in
\cref{sec:conclusions}.

\section{Related Work}\label{sec:related-work}

We categorise related work into two areas: blockchain-based
decentralised governance and token-based voting.

Blockchains provide a fertile testing ground for decentralised
governance.  \citet{schneider-22} discusses the use of economic
incentives, enforced through cryptography and blockchains, to guide
user behaviour, and to impose limitations on the possibilities for
blockchain-based governance.  \citet{reijers-et-al-18} consider the
differences between on-chain and off-chain governance in
blockchain-based systems: they equate on-chain governance to a
mechanical, content-independent understanding of the law.
\citet{dupont-17} provides a historical account of the rise and fall
of The DAO, a \gls*{dao} that was deployed to the Ethereum blockchain
in April 2016.  It is considered a notable failure of blockchain-based
governance.  \citet{bhambhwani-22} examines the governance structure
of the most popular blockchain-based finance applications and
categorises them based on their level of insider control.

In terms of token-based voting, \citet{abramowicz-20} describes formal
games that incentivise players to provide quantitative assessments on
matters of opinion that match the assessments of future players,
thereby eliciting group answers to normative questions.
\citet{merrill-et-al-20} propose a governance model called
\emph{ping-pong governance} where participants lock and, optionally
spend, tokens in order to champion proposals and cast votes.  If a
proposal achieves enough votes within a fixed period, it enters a
review period where participants can try to veto the proposal.  If a
proposal is vetoed, it enters another review period, where
participants can try to veto the veto.  This process is repeated until
the proposal is settled.  The model is superficially similar to the
veToken model discussed in this paper.

\citet{buterin-21} warns against the dangers of token-based voting, in
particular, inequalities and the misalignment of incentives.
\citet{fritsch-et-al-22} analyse the token holders, delegates and
proposals of Compound, Uniswap and \glstext*{ens} over a \num{27}-month period.
They calculate the distribution of voting rights among delegates using
the Gini and Nakamoto coefficients~\citep{srinivasan-lee-17}, the
participation of tokens and delegates in past proposals, and the
potential and exercised voting power of delegates.  In all cases, the
calculations show the existence of a small number of powerful actors.
They classify delegates based on the number of tokens and token
holders that delegate to them.  They show that their voting behaviour
in all cases is similar.  Finally, they perform some rudimentary
address clustering that suggests even more
centralisation~\citep{amico-21}.  \citet{sun-et-al-22-1} analyse voting
behaviour within Sky~\citep{sky-xx} to identify voting coalitions, to
examine their group cohesion and internal structure, and to evaluate
their influence on the protocol.  They draw many parallels between
voting in corporate finance and voting within decentralised
organisations.

\section{The veToken Model}\label{sec:vetoken-model}

The veToken model was proposed and implemented by Michael Egorov as
part of Curve, an automated market maker that specialises in
stablecoin trading~\citep{egorov-20,curve-xx}.  The model was
implemented using smart contracts that extend
Aragon~\citep{cuende-izquierdo-17}, a \gls*{dao} governance framework.
The critical difference between it and previous systems was that it
replaced the one-token-one-vote model with a voting weight
proportional to lock time.

Users need to lock or escrow tokens for a fixed period in exchange for
voting weight.  The longer the period, the more voting weight is
granted.  For Curve, the token is known as \unit{\crv} and the
corresponding locked tokens or voting weight is known as
\unit{\vecrv}\@.  During the lockup period, the locked tokens are
non-transferable, that is, \unit{\crv} is transferable but
\unit{\vecrv} is not.

Voting weight, $w$, for a user can be calculated using:
\begin{equation*}
  w \coloneqq a \cdot \frac{t}{t_{\text{max}}}
\end{equation*}

where $a$ is the number of tokens (for example, \unit{\crv}), $t$ is the
desired lockup period, and $t_{\text{max}}$ is the maximum lockup
period~\citep{egorov-20}.  In Curve, the minimum lockup period is one
week and the maximum is four years.  Consequently, a user who locks up
their tokens for one week would need to lock up \num{208} times the
number of tokens compared to a user who locks for the maximum lockup
period of four years to get the same voting weight.  The intention is
to align voters' potentially short-term interests with the protocol's
long-term goals by tying voting weight to the duration of the lockup.

The veToken model also addresses voter apathy.  Decentralised
organisations typically suffer from poor voter
turnout~\citep{fritsch-et-al-22}, and voting weight alone might not be
incentive enough for users to lock tokens for extended periods.  The
model therefore includes an economic incentive: the allocation of new
tokens (\unit{\crv} tokens in the case of Curve) is directed via
voting using a mechanism known as \emph{gauges}.

\subsection{Gauges}\label{sec:gauges}

Curve is a platform for stablecoin trading.  It uses gauges to measure
the amount of liquidity provided by users to the various liquidity
pools.  Each gauge is assigned a weight and those weights determine
the daily emission of new \unit{\crv} tokens.  The weights are set
every week at Thursday midnight UTC after a gauge proposal vote.
Therefore, users with voting weight decide on the allocation of new
\unit{\crv} tokens to liquidity providers.  In fact, the voters may be
liquidity providers and they can choose to direct the new tokens to
themselves.  This incentivises the lockup of tokens for extended
periods.

\subsection{Higher-Level Protocols}\label{sec:higher-level-protocols}

Curve has spawned several higher-level protocols and strategies.  For
example, yield aggregators such as Convex
Finance~\citep{convex-finance-xx}, Yearn
Finance~\citep{yearn-finance-xx}, and Stake DAO~\citep{stake-dao-xx}
provide smart contracts that hold and lock \unit{\crv} on behalf of
users.  This allows users to lock tokens for the maximum lockup
period, to trade tokenised claims on those locked tokens, and to
amortise gas costs.  Voting markets such as Votium~\citep{votium-xx}
allow anyone to incentivise particular Curve gauges by rewarding votes
with bribes.  Frax Finance~\citep{frax-xx} is a stablecoin issuer that
interacts with many of these protocols.  For example, it uses Curve to
lock \unit{\crv} for \unit{\vecrv}, it holds tokenised \unit{\vecrv}
via Convex, and it incentivises Curve gauges using Votium.

\Cref{fig:dao-governance-curve-convex} illustrates the Curve/Convex
governance ecosystem.  Convex has its own governance token,
\unit{\cvx}, that users lock for voting weight in the form of
\unit{\vlcvx}\@.  Two communities of actors hold the \unit{\vlcvx}
token and vote in Convex, which controls a voter address that holds
\unit{\vecrv} and casts votes in Curve.  Two tokenisation chains
capture the token relationships: \unit{\crv} is tokenised by
\unit{\vecrv} and \unit{\cvx} is tokenised by \unit{\vlcvx}\@.  The
diagram shows how governance decisions at the Convex level propagate
down to Curve through the voter address that connects the two tiers.

\begin{figure}
  \centerline{\myincludegraphics[width=0.7\linewidth]{tikz/dao-governance/dao-governance-curve-convex}}
  \caption{Tokenised governance in the Curve/Convex ecosystem.  Two
    communities hold \unit{\vlcvx} and vote in Convex.  Convex
    controls a voter address that holds \unit{\vecrv} and votes in
    Curve.  \unit{\crv} is tokenised by \unit{\vecrv} and \unit{\cvx}
    is tokenised by \unit{\vlcvx}, creating two layers of governance
    above Curve.}\label{fig:dao-governance-curve-convex}
\end{figure}

\section{Method}\label{sec:method}

We collected on-chain and off-chain data from three primary sources
within the Curve ecosystem.  Firstly, we gathered transaction data
from the Ethereum blockchain relating to Curve: transactions involving
the \unit{\crv} token contract, the \unit{\vecrv} token contract, and
the Gauge Controller contract.  Curve executes its governance entirely
on-chain.  Our data for Curve covers the period from the deployment of
the \unit{\crv} token on 12th August 2020 until 17th March 2023.
Secondly, we gathered transaction data from the Ethereum blockchain
relating to Convex: transactions involving the \unit{\cvx} token
contract, the \unit{\vlcvx} token contract, and the \unit{\cvxcrv}
token contract.  \unit{\cvx} is Convex's governance token;
\unit{\vlcvx} is a locked version of \unit{\cvx} that works in a
manner similar to \unit{\crv} and \unit{\vecrv}, except that the
maximum lockup period is much shorter (\num{16} weeks) and the voting
weight per \unit{\vlcvx} is determined by the amount of \unit{\vecrv}
the Convex protocol owns.  Users that lock \unit{\cvx} to get
\unit{\vlcvx} receive voting weight on Convex, and in turn, voting
weight from the \unit{\vecrv} held by Convex.  Our data for Convex
covers the period from the deployment of the \unit{\cvx} token on 17th
May 2021 until 17th March 2023.  We also collected the details of all
Convex proposals via Snapshot~\citep{snapshot-xx} using its GraphQL
\glstext*{api}\@.  Thirdly, we gathered transaction data from the
Ethereum blockchain relating to Votium: transactions involving the
bribing contracts for the gauges.  Our data for Votium covers the
period from the deployment of the Votium bribing contracts on 12th
September 2021 until 17th March 2023.  We mapped the bribes in the
Votium contracts to their associated gauge in the Curve Gauge
Controller contract.

\section{Results}\label{sec:results}

We begin with results relating to Curve's implementation of the
veToken model.  These can be considered first-order results, in that
they relate directly to \unit{\crv} and \unit{\vecrv}\@.  In the
subsequent subsections
(\cref{sec:votes-follow-bribes,sec:the-cost-of-votes}) we consider two
significant second-order results that relate to higher-level
protocols, namely Convex, Votium, and Frax, that build upon the Curve
ecosystem.

Firstly, we analysed participation in the weekly gauge proposals
and the non-gauge proposals.  The former are economically incentivised
through bribes and \unit{\crv} emissions whereas the latter are not.
We found \num{9551} unique addresses that lock their \unit{\crv}
tokens to get \unit{\vecrv}, with \num{2555} (\qty{27}{\percent})
voting in the gauge proposals.  Participation in
the non-gauge proposals is significantly lower.  For example,
\textquote{ownership proposals}, which seek the community's decision
on adding new gauges, receive votes from an average of \num{24} unique
addresses.  This mirrors the low participation found in other
decentralised organisations~\citep{fritsch-et-al-22} and shows that
the better turnout of economically incentivised gauge proposals does
not carry over to non-gauge proposals.

Secondly, we considered the quantity of \unit{\crv} tokens that are
locked for \unit{\vecrv}: there are \num{644} million \unit{\crv}
(\qty{46}{\percent}) locked for \unit{\vecrv}, leaving \num{770}
million \unit{\crv} in circulation.  The tokens are locked with an
average remaining lockup period of \num{181} weeks, or \num{3.5}
years.  This duration indicates a high level of commitment by the
token holders to the ecosystem and a long-term outlook on the value of
the protocol.  These numbers show the veToken model is working as
expected: users are incentivised to lock up their \unit{\crv} tokens
for extended periods to gain voting weight, and they use that weight
to vote on gauge proposals and direct \unit{\crv} emissions to
liquidity providers.  This enthusiasm does not extend to non-gauge
proposals, where average participation remains at the \num{24}
addresses noted above.

\subsection{Bribes and Voting Outcomes}\label{sec:votes-follow-bribes}

\begin{figure}
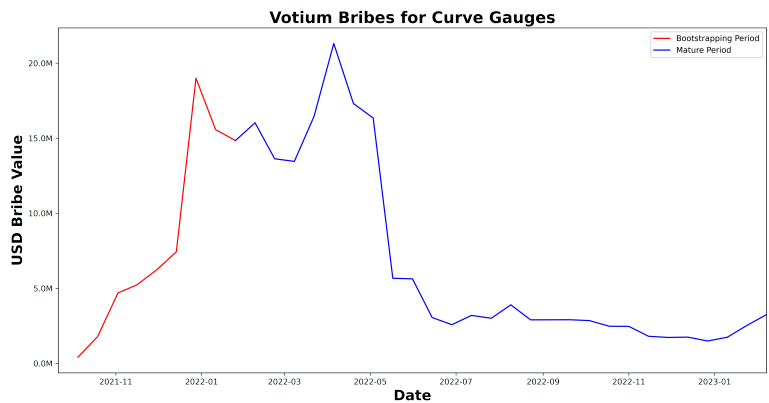

  \centerline{\myincludegraphics[width=\linewidth]{png/figures/curve-votium-bribes}}
  \caption{Votium incentivises holders of \unit{\vecrv} and
    \unit{\vlcvx} to vote for particular Curve gauges through bribes.
    The chart shows the total USD value of those bribes for each
    fortnightly vote.  The USD value is highly dependent on broader
    market conditions.  We divide the data into a bootstrapping phase
  (red) and a mature phase (blue).}\label{fig:votium-bribes}
\end{figure}

\begin{figure}
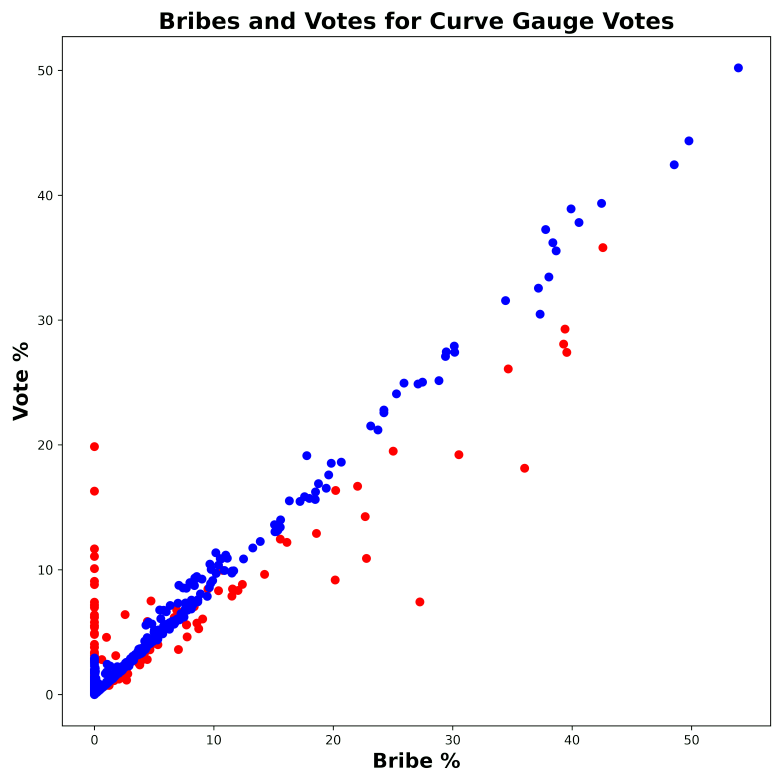

  \centerline{\myincludegraphics[width=\linewidth]{png/figures/curve-bribe-vote-correlation}}
  \caption{Each dot represents a single instance of a fortnightly vote
    for a Curve gauge.  The x-axis shows the percentage of the total
    USD value of Votium bribes directed to each gauge.  The y-axis
    shows the percentage of the total vote received by each gauge.
    The red dots correspond to the bootstrapping phase
    ($\textrm{corr.~coeff.} = 0.88$); the blue dots correspond to the
    mature phase
    ($\textrm{corr.~coeff.} = 0.99$).}\label{fig:votium-bribe-vote-correlation}
\end{figure}

\begin{sidewaysfigure}
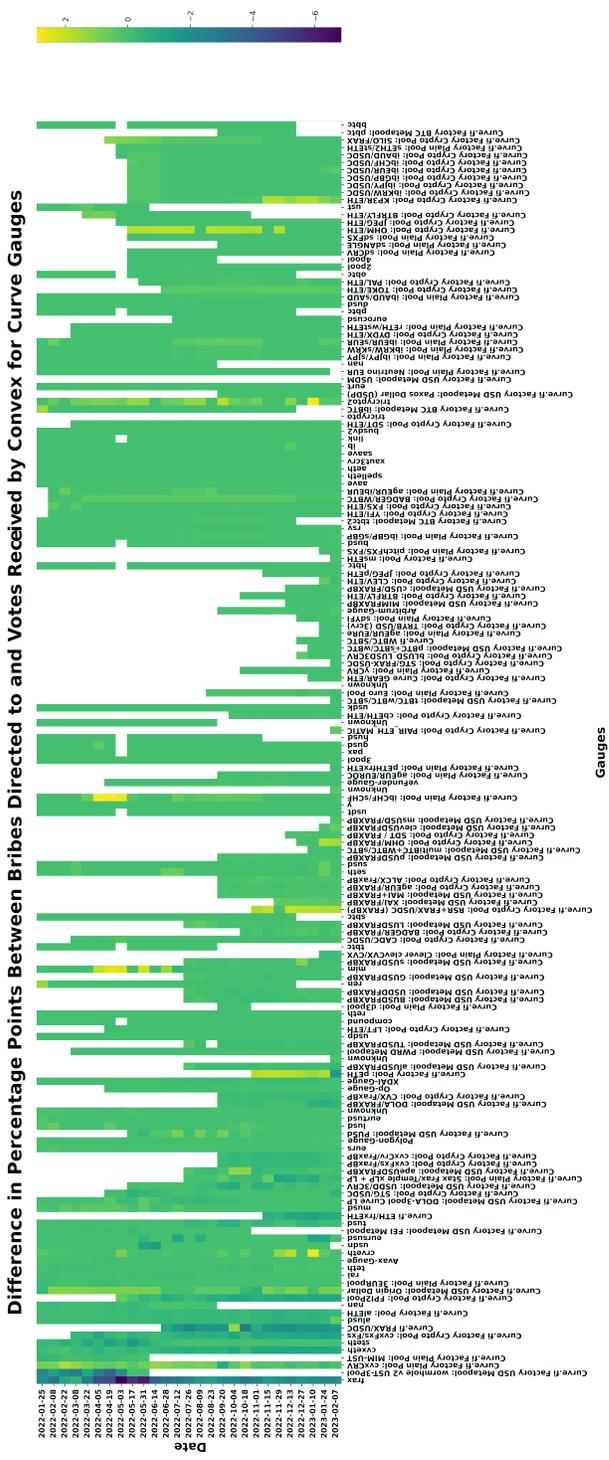

  \centerline{\myincludegraphics[width=\linewidth]{png/figures/curve-votes-follow-bribes}}
  \caption{The non-white cells in the heatmap show the difference in
    percentage points between the bribes directed to and the votes
    received by a Curve gauge for a single instance of a fortnightly
    vote.  The white cells indicate no bribes or votes.  The gauges
    are arranged left-to-right based on the total bribe amount per
    gauge, starting with the
  largest.}\label{fig:votium-votes-follow-bribes}
\end{sidewaysfigure}

We turn our attention to the higher-level protocols.  Convex
participates in the Curve ecosystem as one of nine whitelisted
contract accounts that can lock \unit{\crv} for \unit{\vecrv}\@.  In
fact, \num{290} million (\qty{45}{\percent}) of all locked \unit{\crv}
belongs to Convex, making it the entity with the largest voting
weight in Curve.  However, Convex's voting weight is itself governed
by its own token, \unit{\cvx}\@.  There are \num{71} million
\unit{\cvx}, of which \qty{78}{\percent} is locked for \unit{\vlcvx}
to gain voting weight.  This is higher than the equivalent figure for
Curve.  Voting on Convex is conducted off-chain via Snapshot, with the
results relating to Curve proposals posted on-chain.

As with Curve, the fortnightly gauge proposals attract the most
interest: \num{1124} unique addresses and \qty{95}{\percent} of all
\unit{\vlcvx} vote in the gauge proposals.  The number of unique
addresses might be higher except for delegation, where a single address
can vote using voting weight delegated to it from many other
addresses.  There is less interest in non-gauge proposals, which
receive votes from an average of \num{77} unique addresses.  All of
this voting activity appears as a single address when analysing Curve
in isolation.

Votium operates at a higher level than Convex.  It gathers and
distributes bribes to holders of \unit{\vecrv} and \unit{\vlcvx} in
return for them voting a particular way in the gauge proposals.
\Cref{fig:votium-bribes} shows the total USD value of those bribes for
each fortnightly vote.  The USD value is highly dependent on broader
market conditions: the peaks in late 2021 and early 2022 coincide with
peaks across cryptocurrency markets.  At the time of writing, the
total value of Votium bribes to date is \dUSD[prefix=million~]{248}
and the value for the most recent fortnightly period
is over \dUSD[prefix=million~]{3}.

We can divide the data in \cref{fig:votium-bribes} into a
bootstrapping phase and a mature phase.  The bootstrapping phase
includes the first eight gauge proposals; the mature phase includes
all subsequent gauge proposals and continues to the present day.
\Cref{fig:votium-bribe-vote-correlation} illustrates the change in
voting behaviour between the two phases.  Each dot represents a single
instance of a fortnightly vote for a gauge proposal.  The x-axis shows
the percentage of the total USD value of Votium bribes directed to
each gauge.  The y-axis shows the percentage of the total vote
received by each gauge.  In the bootstrapping phase (red dots) the
correlation coefficient between the percentage of bribes attracted and
the percentage of votes received is \num{0.88}.  In the mature phase
(blue dots) the correlation coefficient is \num{0.99}.

There were few outliers in the mature phase.  Of the \num{691} dots in
\cref{fig:votium-bribe-vote-correlation}, only one represented a gauge
that received a percentage of votes less than \num{0.8} times its
relative bribe in a given fortnight.  In other words, voting outcomes
are strongly associated with bribes.  Conversely, there were \num{91}
instances where a gauge received a percentage of votes more than
\num{1.2} times its relative bribe.  Two overlapping factors account
for these.  Firstly, in \num{71} of these instances the gauge received
a vote that was less than \qty{1}{\percent} of the total number of
votes for a given fortnight; the bribes and votes associated with
these gauges were too small to be meaningful.  Secondly, \num{79} of
these instances involved Frax gauges: since Frax owns \unit{\vlcvx}
and is a major briber to Votium (see \cref{sec:the-cost-of-votes}), it
is to be expected that Frax would vote for its own lesser-bribed
gauges.

A correlation of \num{0.99} demonstrates a strong association between
the distribution of Votium bribes and voting outcomes; it does not, by
itself, establish that bribes directly cause those outcomes.  Votium
is specifically designed to incentivise those with voting power to
direct their voting weight towards particular gauges or to delegate
their voting power to the Votium address, where it will
\textquote{choose the best incentives on every
proposal}~\citep{votium-xx}.  This incentivisation mechanism, combined
with the observed correlation, makes an influence on voting behaviour
plausible; however, other factors may also contribute to the observed
association.  For example, an address with voting power may have
exposure to a gauge and be a beneficiary of emissions from that gauge,
posing an alternative economic incentive, and bribes may anticipate
votes that would have been cast regardless.

This analysis shows a strong association between the monetary value of
the distribution of bribes and the outcome of gauge proposals within
Convex: a gauge that attracts a greater value of bribes generally
receives a greater proportion of votes, suggesting that bribes shape
gauge proposal outcomes.  \Cref{fig:votium-votes-follow-bribes}
illustrates this on a per gauge/gauge proposal basis.  In the heatmap,
the non-white cells show the difference in percentage points between
the bribes directed to and the votes received by a Curve gauge for a
single instance of a fortnightly vote.  The white cells indicate no
bribes or votes.  All of the percentage-point differences fall between
\qty{-6}{\percent} and \qty{2}{\percent}.  As Charlie Munger once
said, \textquote{Show me the incentive, and I will show you the
outcome.}

\subsection{The Cost of Votes}\label{sec:the-cost-of-votes}

\begin{figure}
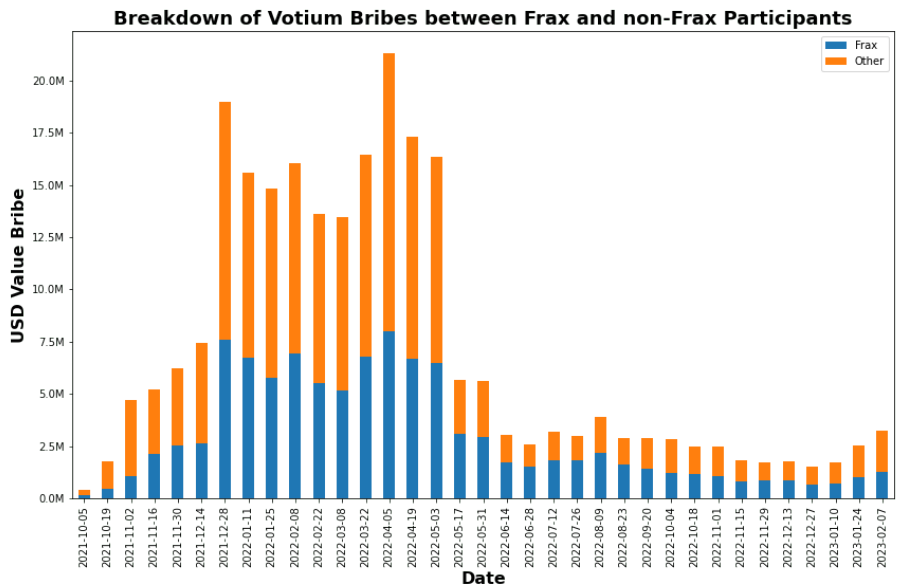

  \centerline{\myincludegraphics[width=\linewidth]{png/figures/curve-votium-frax-bribes}}
  \caption{Frax incentivises holders of \unit{\vecrv} and
    \unit{\vlcvx} to vote for Frax pools through Votium.  It
    contributes almost half of the bribes to
  Votium.}\label{fig:frax-bribes}
\end{figure}

\begin{figure}
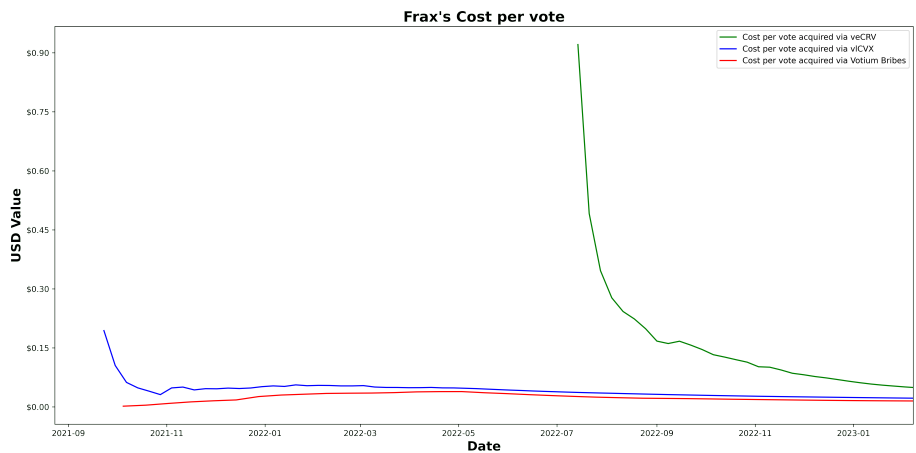

  \centerline{\myincludegraphics[width=\linewidth]{png/figures/curve-frax-cost-per-vote}}
  \caption{Frax acquires votes for Frax pools through at least three
    distinct avenues: by locking \unit{\crv} for \unit{\vecrv}, by
    locking \unit{\cvx} for \unit{\vlcvx}, and by paying Votium
    bribes.  The cost per vote varies depending on the avenue, and the
  cost changes over time.}\label{fig:frax-cost-per-vote}
\end{figure}

Frax is a central player in the Curve ecosystem.  It participates at
the three levels already mentioned: it locks \unit{\crv} for
\unit{\vecrv}, it locks \unit{\cvx} for \unit{\vlcvx}, and it makes
bribes via Votium.  The \unit[per-mode=symbol]{\frax\per\usdc} pool
has the second-highest total value locked across all Curve
pools, behind only the
\unit[per-mode=symbol]{\eth\per\relax~Staked~\eth} pool.  We
considered Frax's involvement in the Curve ecosystem at each level.

At the Curve level, Frax began locking \unit{\crv} for \unit{\vecrv}
in July 2022 after a successful governance proposal whitelisted the
\unit{\frax} staking contract.  As of February 2023, Frax has locked
\num{697114} \unit{\crv} for \unit{\vecrv}, or just
\qty{0.1}{\percent} of all locked \unit{\vecrv}\@.  We determined the
USD value of the locked \unit{\crv} by the price at which it traded
when it was locked.  This resulted in a total locked value of
\dUSD{618830}.  Frax used this voting weight to exercise a total of
\num{17.3} million votes.  This yielded a cost per
vote of \dUSD[base=3]{0.051}.  The cost per vote decreases over time
as the upfront cost of the token is
amortised (see the green line in \cref{fig:frax-cost-per-vote}).
These tokens also have the potential to be traded once the lockup
period has expired.

At the Convex level, Frax plays a more influential role.  Using the
same metric as above, we found that, as of February 2023, Frax has
locked \dUSD[prefix=million~]{64.74} worth of \unit{\cvx} for
\unit{\vlcvx}\@.  To determine its voting weight at the Curve level,
we examined the amount of \unit{\vecrv} tokens held by Convex during
each fortnightly period and the percentage of voting weight held by
Frax.  This showed that Frax exercised a total of \num{2.88} billion
votes, which yields a cost per vote of \dUSD[base=3]{0.022} (see the
blue line in \cref{fig:frax-cost-per-vote}).

Finally, we analysed Frax's bribes via Votium.  Frax is the largest
contributor to bribes, accounting for \qty{41}{\percent} of all bribes
to \unit{\vlcvx}, or \dUSD[prefix=million~]{103.69}
(see \cref{fig:frax-bribes}).  Based on the bribe-vote
correlation reported in \cref{sec:votes-follow-bribes}, we estimate
that this contribution corresponds to \num{6.72} billion votes at the
Curve level.
As of February 2023, Votium bribes for Frax cost \dUSD[base=3]{0.015}
per vote (see the red line in \cref{fig:frax-cost-per-vote}), making
it the most cost-efficient option for Frax.  Unlike the previously
described levels, the value of
bribes is realised immediately, whereas locking tokens still provides
Frax with a claim to the asset once the lock reaches maturity.

The veToken model creates a system where voting weight can be acquired
through three distinct mechanisms.  Most directly, users lock tokens
(say, \unit{\crv} for \unit{\vecrv}) and receive the corresponding
voting weight.  Through an intermediary, users hold a claim to locked
tokens and the voting weight they confer (say, \unit{\cvx} locked for
\unit{\vlcvx}).  At a further remove, users bribe the holders of
locked tokens (\unit{\vecrv} or \unit{\vlcvx} holders) to vote a
particular way.  We can calculate a cost per vote at each of these
levels.

\section{Conclusions}\label{sec:conclusions}

We described the veToken model and its implementation within the Curve
ecosystem.  We analysed its usage by looking at locking rates, lockup
durations, voting participation, and financial incentives.  The
veToken model is intended to align voter and protocol incentives by
tying voting weight to long lockups.  However, through multiple layers
of higher-level protocols, the mechanics of the model can be
distorted.  At higher tiers, the rules of engagement change: voters
can be directed through bribes, and the distribution of those bribes
is strongly associated with the outcome of gauge proposals.  This
effectively replaces four-year lockup periods with fortnightly gauge
proposals in which the strong association between bribes and voting
outcomes is consistent with economic influence on voter behaviour.  We
also showed that the cost per vote depends on how the vote is
acquired: directly by locking at the base level, indirectly through
yield aggregators that lock on behalf of depositors, or through bribes
paid to voting markets.

Our findings show a marked difference in participation between gauge
and non-gauge proposals.  For non-gauge proposals, voter participation
is low, mirroring many other decentralised
organisations~\citep{fritsch-et-al-22}.  For gauge proposals,
participation is significantly higher.  This affirms the design of the
veToken model: users commit to locking tokens for extended durations
in exchange for voting weight, and in turn participate when that
weight has economic consequence.

Participation in gauge proposals is economically incentivised.  Voters
use their voting weight to direct emissions of \unit{\crv}\@.
Higher-level protocols, including yield aggregators such as Convex and
voting markets such as Votium, have emerged to maximise the return on
this weight for users.  Convex's \qty{45}{\percent} share of all
locked \unit{\crv} and the \dUSD[prefix=million~]{248} worth of bribes
routed through Votium are associated with substantial voting power and
economically incentivised voting within the ecosystem: since January
2022, when Votium began to gain recognition, we observed a \num{0.99}
correlation between voting percentages and the share of bribes
distributed to each gauge.  This contrasts with prior accounts of
decentralised governance in which voting coalitions and influential
leaders play the dominant role~\citep{sun-et-al-22-1}.  Protocols
operating at higher tiers of the governance ecosystem can therefore
alter the intended mechanisms of an underlying organisation.  By
changing the rules of engagement at upper tiers, they create more
economically incentivised governance, which in turn accrues voting
power at those tiers.

One entity that bribes through Votium and exerts influence at every
level of the Curve ecosystem is Frax, a stablecoin issuer.  Frax holds
less than \qty{0.1}{\percent} of all \unit{\vecrv}, yet wields
significant influence at higher levels through Convex and through more
than \dUSD[prefix=million~]{100} in Votium bribes.  In this way, Frax
can seek to influence gauge proposal outcomes by incentivising voters
without locking tokens for extended periods.  The maximum lockup
period for Convex is just \num{16} weeks, against four years for
Curve, and Votium imposes no lockup at all.  One might expect such
votes to carry a premium, since they avoid the illiquidity of a long
lock.  The opposite holds: the cost per vote acquired through Convex
is currently lower than the cost through Curve, and the cost through
Votium is lower still.

While our analysis is specific to the Curve-Convex-Votium ecosystem,
our findings hold broader implications for veToken models that use
gauge emissions.  Indeed, similar governance ecosystems exist among
decentralised organisations, with protocols such as Balancer,
Velodrome, and even Frax using veToken governance models with
economically incentivised governance bribes.  While the empirical
findings presented here are specific to Curve, the governance
mechanisms identified apply to other veToken ecosystems that employ
vote-escrow governance, gauge emissions, and economically incentivised
voting.  Moreover, the tokenisation and composition of governance
rights extends beyond veToken models.  For example, protocols such as
Index Coop use composability to wrap governance tokens, creating
additional governance layers.  However, we would expect the dynamics
within these protocols to differ as they do not employ gauge-based
emissions or external vote-incentive markets that economically reward
particular governance outcomes.

In summary, our study shows the dynamic interplay between the levels
of the Curve ecosystem and, more generally, the veToken model.  The
token-locking mechanism is straightforward to understand, but the
higher-level protocols built on top of it produce complex outcomes.
The model was developed to address the limitations of the
one-token-one-vote design.  In doing so it has spawned additional
protocols that capitalise on its behaviour.  In DeFi, this
composability is viewed as a positive (the \textquote{money legos}
metaphor).  For governance, it means that higher-level protocols can
have surprising and undue influence on lower-level ones.  In future
work, we will extend this analysis to other implementations and
adaptations of the veToken model.

\section*{Acknowledgements}

The authors are grateful to the anonymous reviewers that provided
feedback on earlier versions of this article.

\bibliographystyle{my-abbrvnat}
\bibliography{ve-token}

\end{document}